\documentclass[twocolumn,showpacs,preprintnumbers,amsmath,amssymb]{revtex4}
\usepackage{amsmath,amsfonts,latexsym,amssymb,graphics,epsfig,subfigure,color,makeidx}
\usepackage{xcolor}
\usepackage[colorlinks,
            linkcolor=blue,
            anchorcolor=blue,
            citecolor=green
            ]{hyperref}

\begin{document}

\title{Spinning test particle in four-dimensional Einstein-Gauss-Bonnet Black Hole}

\author{Yu-Peng Zhang\footnote{zhangyupeng14@lzu.edu.cn},
       Shao-Wen Wei\footnote{weishw@lzu.edu.cn},
       Yu-Xiao Liu \footnote{liuyx@lzu.edu.cn, corresponding author}
}
\affiliation{Institute of Theoretical Physics \& Research Center of Gravitation, Lanzhou University, Lanzhou 730000, China\\
Key Laboratory for Magnetism and Magnetic of the Ministry of Education, Lanzhou University, Lanzhou 730000, China\\
Joint Research Center for Physics, Lanzhou University and Qinghai Normal University, Lanzhou 730000 and Xining 810000, China}

\begin{abstract}
In this paper, we investigate the motion of a classical spinning test particle orbiting around a static spherically symmetric black hole in a novel four-dimensional Einstein-Gauss-Bonnet gravity [D. Glavan and C. Lin, Phys. Rev. Lett. 124, 081301 (2020)]. We find that the effective potential of a spinning test particle in the background of the black hole has two minima when the Gauss-Bonnet coupling parameter $\alpha$ is nearly in a special range $-6.1<\alpha/M^2<-2$ ($M$ is the mass of the black hole), which means such particle can be in two separate orbits with the same spin angular momentum and orbital angular momentum. We also investigate the innermost stable circular orbits of the spinning test particle and find that the effect of the particle spin on the the innermost stable circular is similar to the case of the four-dimensional black hole in general relativity.
\end{abstract}

\maketitle

\section{Introduction}\label{scheme1}

As the most successful gravitational theory, general relativity (GR) can explain the relation between geometry and matter. {One of} the most impressive result derived from GR is black hole solutions. As vacuum solutions of strong gravity systems, black holes have lots of interesting characters, for examples, a binary black hole system can produce gravitational waves \cite{Abbott2016a}, and a black hole can act as an accelerator of particles \cite{Banados2009,Wei2009}. But we should note that, even GR is so powerful and can be used to explain much phenomena, there are still some problems that can not be interpreted by GR. Therefore, it is believed that there should be more fundamental {theory beyond GR}.

{It is well known that} the existence of a singularity locating at the inner of {a black} hole {leads to} the geodesics incompleteness \cite{Penrose1965,Hawking1970}. To overcome the problem of singularity, several quantum theories of gravity have been proposed, like the superstring/M theory and the extension of such theories. With the help of perturbation approximation of these theories, the Gauss-Bonnet (GB) term was found as the next leading order term \cite{Gross1986,Bento1996}, and this term is ghost-free combinations and does not add higher derivative terms into the {gravitational field equations}  \cite{Zwiebach1985}. The GB term appears in $D$-dimensional spacetime as follows
\begin{equation}
S_{[\text{GB}]}[g_{\mu\nu}]=\int{d^Dx\sqrt{-g}\alpha\mathcal{G}},
\end{equation}
where $D$ is the number of the spacetime dimensions, {$\alpha$ is the GB coupling parameter with mass dimension $D-4$}, and the GB invariant $\mathcal{G}$ is defined as
\begin{equation}
\mathcal{G}={R^{\mu\nu\rho\sigma}R_{\mu\nu\rho\sigma}-4R^{\mu\nu} R_{\mu\nu}}+R^2.
\end{equation}
Black holes solutions of GB gravity in $D\geq 5$ have been derived, such as the vacuum case \cite{Boulware1985}, Einstein-Maxwell fields with a GB term \cite{Wiltshire1986,Wiltshire1988}, and anti-de Sitter (AdS) case \cite{Cai2002}. In four-dimensional spacetime, the GB term does not make contributions to the gravitational dynamics, which leads to that the four-dimensional minimally {coupled} GB gravity is hard to get. However, very recently, D. Glavan and C. Lin \cite{Glavan2020} proposed a novel four-dimensional Einstein-Gauss-Bonnet (EGB) gravity that bypasses the Lovelock theorem by adopting an artful coupling constant $\alpha\to \frac{\alpha}{D-4}$. In this novel four-dimensional EGB gravity, the GB invariant term does not affect the properties of the massless graviton and a four-dimensional static and spherically symmetric black hole solution {was obtained}. The stability and shadow of this four-dimensional EGB black hole have been studied in \cite{Konoplya2020}, where the quasinormal modes of a scalar, electromagnetic, and gravitational perturbations {were studied}. {The solutions} of charged black hole \cite{Fernandes2020} {and spinning black hole} \cite{Wei2020} {were also} been obtained, and {constraint} to the {GB parameter $\alpha$} {was first} given in Ref.~\cite{Wei2020} in terms of the shadow of the rotating black hole. Inspired by the novel four-dimensional EGB gravity, the novel four-dimensional Einstein-Lovelock gravities are also proposed \cite{Konoplya:2020qqh,Casalino:2020kbt}.

To understand the geometry of a black hole, the behaviors of the geodesics of a test particle around a black hole can be used. We know that a massless or massive particle can orbit around a central black hole and the motion is dependent on the geometry of the central black hole. In Ref. \cite{Guo2020}, the innermost stable circular orbit (ISCO) of a spinless test particle and shadow in the background of the four-dimensional EGB black hole {were} studied. {The range} of the GB coupling parameter for the black hole solution {was extended} to the range of {$-8\leq \alpha/M^2\leq 1$}, where $M$ is the mass of the black hole. Compared to the case of a test particle in the background of a Schwarzchild black hole in GR, a positive GB coupling parameter {reduces the radii of the ISCOs} and {a negative one enlarges them}. {There also appears Similar effect} for a spinning test particle in the background of a black hole in GR, where the spin-curvature force $-\frac{1}{2}R^\mu_{\nu\alpha\beta}u^\nu S^{\alpha\beta}$ also {reduces or enlarges the corresponding radii of the ISCOs} \cite{Suzuki:1997by}. Inspired by the {effects of the four-dimensional GB term and the non-vanishing spin of a particle on the motion of the test particle}, it is necessary to investigate what the total {effects of them} on the motion of a test particle in the four-dimensional EGB black hole. In this paper, we will investigate the motion of a spinning test particle in the background of the novel four-dimensional EGB black hole. For simplicity, we only consider the motion of {a spinning} test particle in the equatorial plane.

For a spinning test particle, {its motion} will not follow the geodesics because of the spin-curvature force $-\frac{1}{2} { R^\mu_{~\nu\alpha\beta} } u^\nu S^{\alpha\beta}$. The equations of motion for the spinning test particle are described by the Mathisson-Papapetrou-Dixon (MPD) equations \cite{Mathisson,Papapetrou1951a,Papapetrou1951c,Dixon,phdthesis,Hojman1977,Bahram2006,Zalaquett2014,Ruangsri:2015cvg} under the ``pole-dipole'' approximation, and the four-velocity $u^\mu$ and the four-momentum $P^\mu$ are not parallel \cite{phdthesis,Armaza2016} due to the spin-curvature force. The four-momentum $P^\mu$ of a spinning test particle keeps timelike along the trajectory ($P^\mu P_\mu=-m^2$, $m$ is the mass of the test particle), in the contrary, the four-velocity
would be superluminal \cite{phdthesis,Armaza2016} when the spin of the
test particle is too large. Actually, this superluminal behavior comes from the ignorance of the ``multi-pole'' effects. {When} such effects are considered, the superluminal problem can be avoided
\cite{Deriglazov:2015zta,Deriglazov:2015wde,Ramirez:2017pmp,Deriglazov:2017jub,Jan2010}. For the properties of the spinning test {particle} in different black hole backgrounds, see Refs.~\cite{Han2010,Harms:2016ctx,Lukes-Gerakopoulos:2017vkj,Suzuki:1997by,Pugliese:2013zma,zhang2018:zwgsl,Sajal2018,Zhang:2018eau,stuchlik_1999,stuchlik_2006,Plyatsko_2017,Plyatsko_2018,Han2008,Warburton2017,Liu:2018myg,Mukherjee:2018kju,Faye:2006gx,Zhang:2016btg,Conde2019,Lukes-Gerakopoulos2019,Jefremov:2015gza,Toshmatov2019,Nucamendi2020,liu2020}.

This paper is organized as follows. In Sec.~\ref{scheme1}, we use the MPD equation to obtain the four-momentum and four-velocity of {a spinning} test particle in the novel four-dimensional EGB black hole background. In Sec.~\ref{scheme2}, we study the motion of the spinning test particle and give the relations between the motion of {the spinning} test particle and the properties of the four-dimensional EGB black hole. Finally, a brief summary and conclusion are given in Sec.~\ref{Conclusion}.

\section{Motion of a spinning test particle in four-dimensional EGB black hole}{\label{scheme1}}
\subsection{Four-momentum and four-velocity of the spinning test particle}

In this part, we will solve the equations of motion for {a spinning} test particle in the novel four-dimensional EGB black hole background.

The action of the $D$-dimensional EGB gravity is described by{
\begin{eqnarray}
S =  \int{d^D x \sqrt{-g}
  \left[\frac{1}{2\kappa^2}R+\alpha\mathcal{G}\right]},
\end{eqnarray}
where $\kappa$ is the gravitational constant and will be set as $\kappa^2=1/2$ in this paper.} The GB term does not contribute to the dynamics of the four-dimensional spacetime because {it is} a total derivative. {Recently, by rescaling the coupling parameter as
\begin{equation}
\alpha\to \frac{\alpha}{D-4},
\end{equation}
and taking the limit $D\to 4$, Glaan and Lin \cite{Glavan2020} obtained the four-dimensional novel EGB gravity.} {The} four-dimensional static spherically symmetric black hole solution {was found} \cite{Glavan2020}
    \begin{eqnarray}
    ds^2&=&-f(r)dt^2+\frac{dr^2}{f(r)}+r^2d\Omega^2,\label{metric}\\
    f(r)&=&1+\frac{r^2}{2\alpha}\left(1-\sqrt{1+\frac{8\alpha M}{r^3}}\right), \label{fr}
\end{eqnarray}
where $M$ is the mass of the black hole and the coupling parameter $-8\leq \frac{\alpha}{M^2} \leq 1$ \cite{Guo2020}. Solving $f(r)=0$, {one can get two black hole horizons}
\begin{equation}
r_{\pm}=M\pm\sqrt{M^2-\alpha}. \label{horizons}
\end{equation}
{In fact, the above solution (\ref{metric}-\ref{horizons}) was also found in gravity with a conformal anomaly in Ref. \cite{Cai2010} and was extended to the case with a cosmological in Ref. \cite{Cai2014}.}

The motion of a spinning test particle is described by the MPD equations
    \begin{eqnarray}
    \frac{D P^{\mu}}{D \lambda} &=& -\frac{1}{2}R^\mu_{\nu\alpha\beta}u^\nu S^{\alpha\beta},\label{equationmotion1}\\
    \frac{D S^{\mu\nu}}{D \lambda} &=&P^\mu u^\nu-u^\mu P^\nu, \label{equationmotion2}
    \end{eqnarray}
where $P^{\mu}$, $S^{\mu\nu}$, and $u^{\mu}$ are the four-momentum, spin tensor, and tangent vector of the spinning test particle along the trajectory, respectively. Note that the MPD equations are not uniquely specified and we should use a spin-supplementary condition to determine {them}. This spin-supplementary condition is related to the center of mass of the spinning test particle with different observers \cite{Wald:1972sz,Lukes-Gerakopoulos2014,Costa2017,Lukes-Gerakopoulos2017,Georgios2017}. In this paper, we choose the Tulczyjew spin-supplementary condition \cite{Tulczyjew}
\begin{equation}
P_\mu S^{\mu\nu}=0,
\label{supplementarycondition}
\end{equation}
and the four-momentum $P^\mu$ satisfies
\begin{equation}
P^\mu P_\mu=-m^2,\label{normal1}
\end{equation}
which makes sure that the spinning test particle keeps timelike along the trajectory.

For the equatorial motion of the spinning test particle with spin-aligned or anti-aligned orbits, the four-momentum and spin tensor should satisfy $P^\theta=0$ and $S^{\theta \mu}=0$. The non-vanishing independent variables for the equatorial orbits are $P^t$, $P^r$, $P^\phi$, and $S^{r\phi}$. After adopting the spin-supplementary condition (\ref{supplementarycondition}), we have \cite{Hojman1977}
\begin{eqnarray}
S^{r t}=-S^{r\phi}\frac{P_{\phi}}{P_t}, ~~~~S^{\phi t}=S^{r\phi}\frac{P_r}{P_t}.
\label{spintensor}
\end{eqnarray}
Substituting Eq. (\ref{spintensor}) into {the following} equation
\begin{equation}
s^2=\frac{1}{2}S^{\mu\nu}S_{\mu\nu}=S^{\phi r}S_{\phi r}+S^{t r}S_{t r}+S^{t\phi}S_{t\phi},
\end{equation}
{and using} Eq. (\ref{normal1}), we get the $r-\phi$ component of spin tensor
\begin{equation}
S^{r\phi}=-\frac{s}{r}\frac{P_{t}}{m}.
\end{equation}
The non-vanishing components of the spin tensor $S^{\mu\nu}$ in the four-dimensional EGB black hole background are
\begin{eqnarray}
S^{r\phi}&=&-S^{\phi r}=-\frac{s}{r}\frac{P_{t}}{m},\nonumber\\
S^{rt}&=&-S^{tr}=-S^{r\phi}\frac{P_\phi}{P_t}=\frac{s}{r}\frac{P_\phi}{m},\label{spinnozo}\\
S^{\phi t}&=&-S^{t\phi}=S^{r\phi}\frac{P_r}{P_t}=-\frac{s}{r}\frac{P_r}{m},\nonumber
\end{eqnarray}
where the parameter $s$ is the spin angular momentum of the test particle and the spin direction is perpendicular to the equatorial plane.

Due to the existence of the spin-curvature coupling term, the conserved quantities of the spinning test particle are modified. The {relation} between a killing vector field $\mathcal{K}^\mu$ and the conserved quantity {is} \cite{Hojman1977,phdthesis}
\begin{equation}
\mathcal{C}=\mathcal{K}^\mu P_\mu-\dfrac{1}{2} S^{\mu \nu}\mathcal{K}_{\mu;\nu},
\label{Eq:Conserved_quantity}
\end{equation}
where the semicolon denotes the covariant derivative. { In the EGB black hole with the metric (\ref{metric}), there are two Killing vectors, a timelike $\xi^\mu=(\partial_t)^\mu$ and a spacelike $\eta^\mu=(\partial_\phi)^\mu$, from which we can} get two conserved quantities \cite{Hojman1977}
\begin{eqnarray}
m\bar{e}&=&-\mathcal{C}_t=-\xi^\mu P_\mu+\dfrac{1}{2} S^{\mu \nu}{\xi}_{\mu;\nu}\nonumber\\
&=&-P_t+\frac{1}{2}\frac{\bar{s}}{r}P_t\partial_rg_{t\phi}-\frac{1}{2}\frac{\bar{s}}{r}P_\phi\partial_r g_{tt},\label{conservedenergy}\\
m\bar{j}&=&\mathcal{C}_{\phi}=\eta^\mu P_\mu-\dfrac{1}{2} S^{\mu \nu}{\eta}_{\mu;\nu}\nonumber\\
&=&P_\phi+\frac{1}{2}\frac{\bar{s}}{r}P_\phi\partial_r g_{\phi t}-\frac{1}{2}\frac{\bar{s}}{r}P_t\partial_r g_{\phi\phi}.\label{conservedmomentum}
\end{eqnarray}
{Here} the parameters are defined as $\bar{e}=\frac{e}{m}$, $\bar{j}=\frac{j}{m}$, and $\bar{s}=\frac{s}{m}$, {with} $e$, $m$, and $j$ the energy, mass, and total angular momentum of the spinning test particle, respectively. {Note that we have used} the relations
$S^{\mu \nu}{\xi}_{\mu;\nu}=S^{\mu\nu}\xi^\beta\partial_\nu g_{\beta \mu}$ and $S^{\mu \nu}{\eta}_{\mu;\nu}=S^{\mu\nu}\eta^\beta\partial_\nu g_{\beta \mu}$ for the two Killing vectors.

Solving Eqs. \eqref{normal1}, \eqref{conservedenergy}, and \eqref{conservedmomentum}, we get the non-vanishing  components of the four-momentum:
\begin{eqnarray}
\!\!\!P_t \!\!\!&=&\!\!\!-\frac{m^2 \left(\alpha \left(2 \bar{e} r^3 \Delta+2 \bar{j} M^2 \bar{s}\right)-\bar{j} M r^3 \bar{s} \left(\Delta-1\right)\right)}{\alpha \left(2 r^3 \Delta+2 M \bar{s}^2\right)-r^3 \bar{s}^2 \left(\Delta-1\right)},~~~~~\label{memantumpt}\\
P_\phi \!\!\!&=&\!\!\! \frac{2 \alpha m^2 r^3 \Delta (\bar{j} M-\bar{e} \bar{s})}{\alpha \left(2 r^3 \Delta+2 M \bar{s}^2\right)-r^3 \bar{s}^2 \left(\Delta-1\right)},\label{memantumpp}
\end{eqnarray}
and
\begin{equation}
(P^r)^2 =-\frac{m^2+g^{\phi\phi}P_\phi^2+2g^{\phi t}P_\phi P_t+g^{tt}P_t^2}{g_{rr}},  \label{memantumpr}
\end{equation}
where the function $\Delta=\sqrt{1+\frac{8\alpha M}{r^3}}$. We can solve the four-velocity $u^\mu$ by using the equations of motion (\ref{equationmotion1}) and (\ref{equationmotion2})  and the components of $S^{\mu\nu}$ in (\ref{spinnozo}) \cite{Hojman2013,Zhang:2016btg}
\begin{eqnarray}
\frac{DS^{tr}}{D\lambda} \!\!&=&\!\! P^t\dot{r}-P^r=\frac{\bar{s}}{2r}g_{\phi \mu}R^\mu_{\nu\alpha\beta}u^\nu S^{\alpha\beta}
           +\frac{\bar{s}}{r^2}P_\phi\dot{r}, \label{spinvelocityequation1}\\
\frac{DS^{t\phi}}{D\lambda}
       \!\! &=&\!\!  P^t\dot{\phi}-P^\phi
        = -\frac{\bar{s}}{2r}g_{r \mu}R^\mu_{\nu\alpha\beta}u^\nu S^{\alpha\beta}
            -\frac{\bar{s}}{r^2}P_r\dot{r}.  ~~~~ \label{spinvelocityequation2}.
\end{eqnarray}
Finally, the non-vanishing components of the four-velocity {are obtained as}
\begin{eqnarray}
\dot{r}&=&\frac{b_2 c_1-b_1c_2}{a_2b_1-a_1b_2},\\
\dot{\phi}&=&\frac{a_2 c_1-a_1c_2}{a_1b_2-a_2b_1},
\end{eqnarray}
where the functions $a_1$, $b_1$, $c_1$, $a_2$, $b_2$, and $c_2$ are defined as
\begin{eqnarray}
a_1&=&P^t-\frac{\bar{s}}{r^2}P_\phi+\frac{\bar{s}}{2r}R_{\phi r\mu\nu}S^{\nu\mu},\\
b_1&=&\frac{\bar{s}}{2r}R_{\phi \phi\mu\nu}S^{\nu\mu},\\
c_1&=&-P^r+\frac{\bar{s}}{2r}R_{\phi t\mu\nu}S^{\nu\mu},\\
a_2&=&\frac{\bar{s}P_r}{r^2}-\frac{\bar{s}}{2r}R_{r r\mu\nu}S^{\nu\mu},\\
b_2&=&P^t-\frac{\bar{s}}{2r}R_{r\phi \mu\nu}S^{\nu\mu},\\
c_2&=&-P^\phi-\frac{\bar{s}}{2r}R_{rt\mu\nu}S^{\nu\mu}.
\end{eqnarray}
We can set the affine parameter $\lambda$ as coordinate time and choose $u^t=1$ because the trajectories of the test particle are independent of the affine parameter $\lambda$ \cite{Dixon,Georgios2017}. {Then} the orbital frequency parameter $\Omega$ of the test particle is
\begin{equation}
\Omega\equiv\frac{u^{\phi}}{u^t}=\dot{\phi}.
\end{equation}

\subsection{Properties of a spinning particle in circular orbits}\label{scheme2}

The motion of a test particle in a central field can be solved in terms of the radial coordinate in the Newtonian dynamics \cite{Kaplan,Landau}. We can use the same {way} to simplify the motion of a test particle in the black hole background by using the effective potential method in general relativity. The radial velocity $u^r$ is parallel to the radial momentum $P^r$, therefore the effective potential of the spinning test particle can be solved by using the form of $P^r$ \eqref{memantumpr} \cite{Jefremov:2015gza}. We decompose the $\left(P^r\right)^2$ \eqref{memantumpr} as  \cite{zhang2018:zwgsl,Jefremov:2015gza}
\begin{widetext}
    \begin{eqnarray}
    \frac{(P^r)^2}{m^2}&=&\left(A \bar{e}^2+B \bar{e}+C\right)\propto\left(\bar{e}-\frac{-B+\sqrt{B^2-4AC}}{2A}\right)\left(\bar{e}+\frac{B+\sqrt{B^2-4AC}}{2A}\right),
    \label{effectivepotentiala}
    \end{eqnarray}
\end{widetext}
where the functions $A$, $B$, and $C$ are
\begin{widetext}
\begin{eqnarray}
A &=&2\mathcal{E}^{-1} \alpha m^2 r \left(8 \alpha M+r^3\right) \bigg(r^2 \bar{s}^2(\Delta-1)+2 \alpha \left(r^2-\bar{s}^2\right)\bigg),\\
B &=&8\mathcal{E}^{-1} \alpha^2 \bar{j} m^2 M r \bar{s} \left(-3 M r^2 \Delta+8 \alpha M+r^3\right),
\end{eqnarray}
and
\begin{eqnarray}
C &=& -2 m^2\left(\alpha \mathcal{E}\right)^{-1}
  \Bigg\{16 \alpha^4 M r \left(\bar{j}^2  M^2+r^2\right)
         +\alpha^3 \bigg[2 \bar{j}^2 M^2 \big(4 M r^3 (1-\Delta)-M^2 \bar{s}^2+r^4\big)+4 M r^3 \bar{s}^2 \left(\Delta-4\right)\nonumber\\
        &&+M r^5 \left(8-8 \Delta\right)+2 M^2 \bar{s}^2 \left(\bar{s}^2-8 r^2\right)+2 r^6\bigg]
          +\alpha^2 \bigg[\bar{j}^2 M^2 r^3 \bigg(2 M \bar{s}^2 \left(\Delta-3\right)+r^3 \left(1-\Delta\right)\bigg)\nonumber\\
        &&-M^2 r^2 \bar{s}^4 \left(\Delta-9\right)-2 M r^3 \bar{s}^4 \left(\Delta-3\right)+r^8 \left(1-\Delta\right)+2 r^6
          \bar{s}^2 \left(\Delta-1\right)+2 M r^5 \bar{s}^2 \left(5 \Delta-9\right)\bigg]\nonumber\\
        &&+\alpha r^5 \bar{s}^2 \bigg[\bar{j}^2 M^2 r \left(\Delta-1\right)+r \bar{s}^2 \left(1-\Delta\right)-4 M \bar{s}^2 \left(\Delta-2\right)+r^3 \left(2 \Delta-2\right)\bigg]
         +r^8 \bar{s}^4 \left(\left(1-\Delta\right)\right)
 \Bigg\},\label{gamma1}
\end{eqnarray}
\end{widetext}
where the function $\mathcal{E}$ is
\begin{equation}
\mathcal{E}=\left[r^3 \bar{s}^2 \left(\Delta-1\right)-2\alpha \left( r^3 \Delta+ M \bar{s}^2\right)\right]^2.
\end{equation}
The effective potential of the test particle is defined by the positive square root of Eq. (\ref{effectivepotentiala})
\begin{equation}
V_{\text{eff}}^{\text{spin}}  = \frac{-B+\sqrt{B^2-4AC}}{2A}
\label{spinningeffective}
\end{equation}
{The} positive square root corresponds to the four-momentum pointing toward future, while {the negative one} corresponds to the past-pointing four-momentum \cite{misnerthornewheeler}. When the spin of the test particle is zero, it reduces to
\begin{eqnarray}
V_{\text{eff}}&=&\sqrt{1+\frac{r^2}{2\alpha}\left(1-\sqrt{1+\frac{8\alpha M}{r^3}}\right)}\sqrt{1+\frac{\bar{j}^2}{r^2}}\nonumber\\
&=&\sqrt{f(r)\left(1+\frac{\bar{j}^2}{r^2}\right)}.
\end{eqnarray}
Note that for the four-dimensional Schwarzchild black hole in GR, the function $f(r)=1-\frac{2M}{r}$.

\begin{figure*}[!htb]
\includegraphics[width=\linewidth]{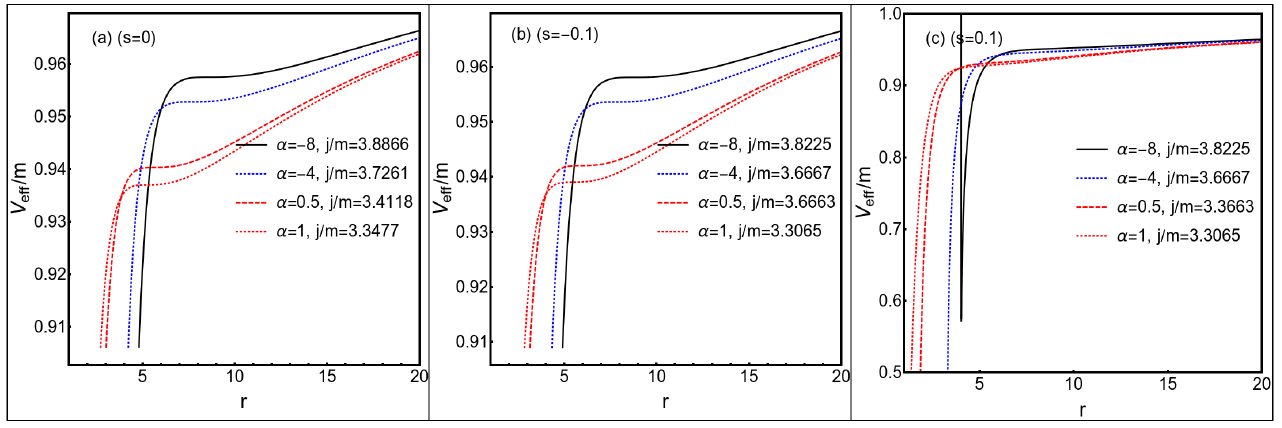}
\caption{{The effective potential for a} spinning test particle in the four-dimensional EGB black hole {background. The} parameters are set as $M=1$ and $m=1$. }
\label{effectivefigure}
\end{figure*}

The properties of a test particle in a central field are mainly determined by the effective potential. {Thus}, the effects on the motion of {a spinning} test particle can be derived based on how the effective potential depends on the GB coupling parameter $\alpha$ and the spin angular momentum $\bar{s}$. We plot some {shapes} of the effective {potential} (\ref{spinningeffective}) in Fig.~\ref{effectivefigure}. We can see that the radii of the extreme points become smaller when the coupling parameter $\alpha>0$ {and become} larger when the parameter $\alpha<0$. These phenomena mean that a positive GB coupling parameter induces the attractive {effect} and a negative {one results in} the repulsive effect on the motion of the test particle.

In addition to {the attractive} or repulsive effects on the motion of {the test} particle, {some} more interesting results are found when we check the shapes of the effective {potential} in the parameter space $(\bar{s}-\bar{j})$. {We find} that the effective {potential has} two minima when the GB coupling parameter $\alpha$ is in a special range. When the test {particle} move in stable circular orbits \cite{Jefremov:2015gza}, the radial velocity should be zero
\begin{equation}
\frac{dr}{d\lambda}=0\label{condition1},
\end{equation}
and the the radial acceleration vanishes
\begin{equation}
\frac{d^2r}{d\lambda^2}=0,~
\left(\frac{dV_{\text{eff}}}{dr}=0~\text{and}~\frac{d^2V_{\text{eff}}}{dr^2}>0\right)\label{condition2}.
\end{equation}
{The} conditions $\frac{dV_{\text{eff}}}{dr}=0$ and $\frac{d^2V_{\text{eff}}}{dr^2}>0$ mean that the energy of {the} particle should equal to the minimum of the effective potential.

\begin{figure*}[!htb]
\includegraphics[width=\linewidth]{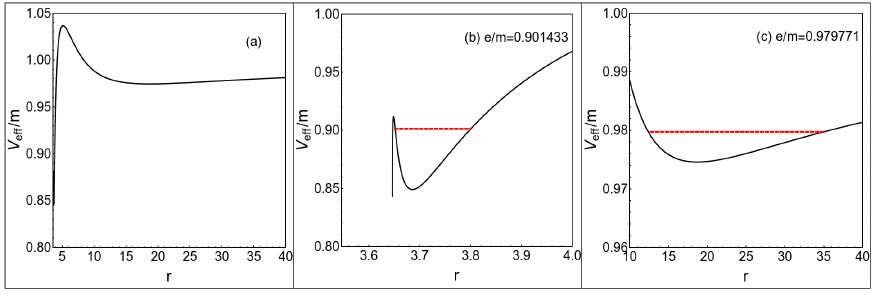}
\includegraphics[width=\linewidth]{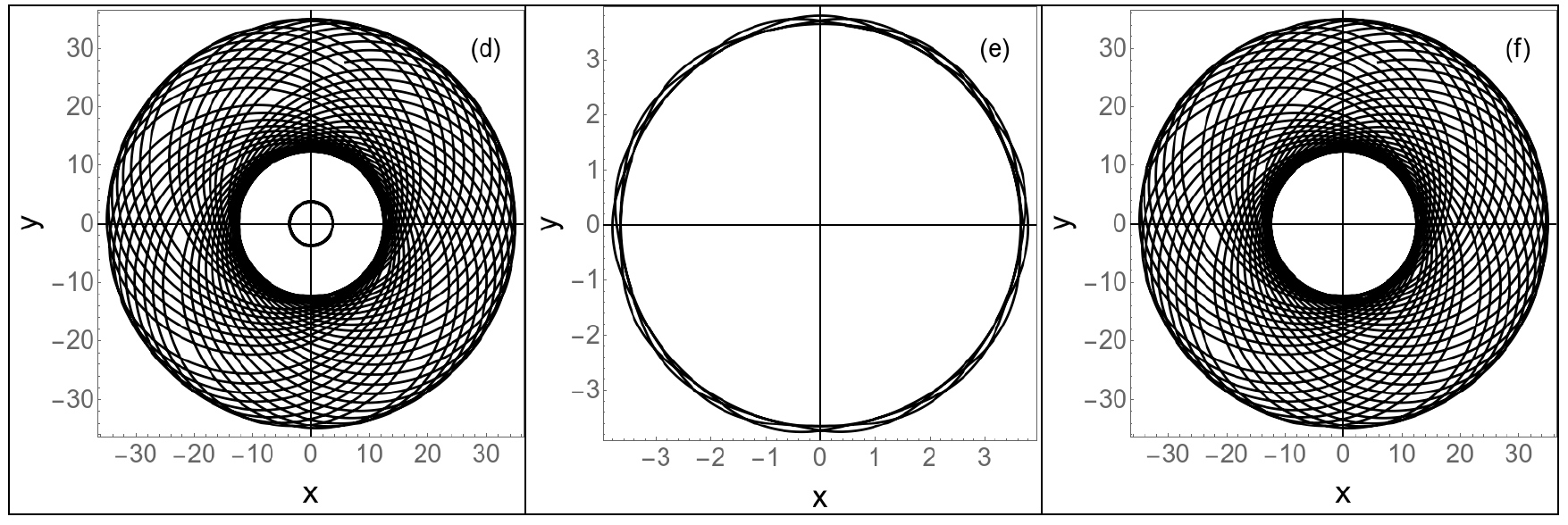}
\caption{Plots of {the} effective potential and orbits for a spinning test particle with $\bar{s}=0.3$, $\bar{j}=5$, and $\alpha=-6$. The subfigures (b) and (c) are the two minima of the effective {potential shown in the subfigure (a)}. The subfigures (e) and (f) are two {separate} orbits around the two minima of the effective potential {shown in the subfigure (d)}, and they are related to the effective potential in the subfigures (b) and (c).  The values of the red dashed line in the subfigures (b) and (c) {stand} for the energy of the test particle. And the range {of the red dashed} in the radial direction stands for {the radial range that the test particle can move in}, see the corresponding orbits in subfigures (e) and (f). The test particles on the two orbits have the same spin and orbital angular momentum. The parameters are set as $M=1$ and $m=1$.
}
\label{potential-orbits}
\end{figure*}

Therefore, when the effective potential of a spinning test particle has two minima, there will be two stable circular orbits for {the particle with a spin angular momentum and an orbital angular momentum.} {This is a} new feature for the motion of a spinning test particle in four-dimensional EGB black hole background. We plot {the effective potential} with two minima in Fig. \ref{potential-orbits}, where the corresponding two separate orbits of the spinning test particle with $\bar{s}=0.3$ and $\bar{j}=5$ are still given.

\begin{figure*}[!htb]
\includegraphics[width=\linewidth]{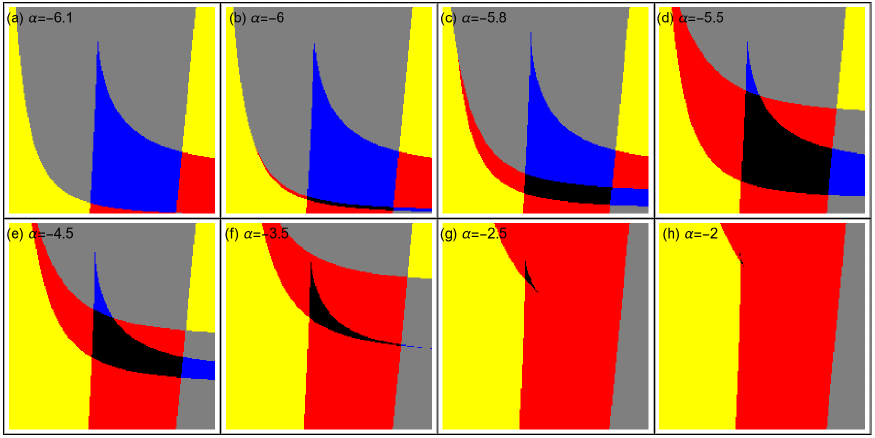}
\caption{Plots of {the parameter space $(\bar{s}-\bar{j})$ describing} whether a spinning test particle can move in two circular orbits with {the same} spin $\bar{s}$ and {the same} total angular momentum $\bar{j}$. {The} vertical axis is the spin parameter in the range of $\bar{s}\in(0.2,1.2)$, and the horizontal axis is the total angular momentum $\bar{j}$ in the range of $\bar{j}\in(0,10)$. The other parameters are set as $M=1$ and $m=1$. {In the black and blue regions, the effective potential has two minima (corresponding to two stable circular orbits) and one minimum (corresponding to one stable circular orbit), respectively. In the gray and yellow regions, the test particle has no stable circular orbits.}}
\label{double-circular-orbits}
\end{figure*}

The spinning test particle with with {the same spin $\bar{s}$ and the same angular momentum} can posses two stable circular orbits only happens in the case of $\alpha<0$ with a special range for $\alpha$. We give the numerical results in Fig. \ref{double-circular-orbits} {and find that} the range of $\alpha/M^2$ is nearly in $(-6.1,-2)$.

\begin{figure*}[!htb]
\includegraphics[width=\linewidth]{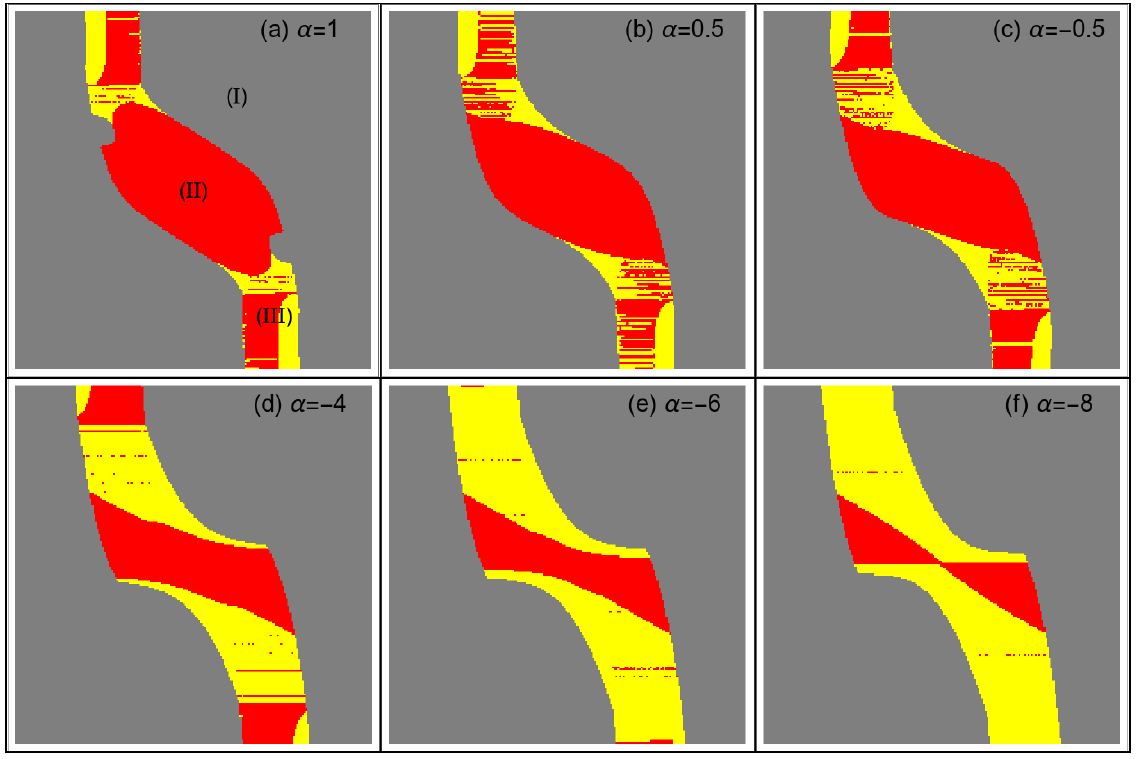}
\caption{Properties of circular orbits for the spinning test particle in the four-dimensional GB black hole background. {The} vertical axis is the spin parameter in the range of $\bar{s}\in(-8,8)$, and {the horizontal} axis is the orbital angular momentum $\bar{l}=\bar{j}-\bar{s}$ in the range of $\bar{l}\in(-8,8)$. The other parameters are set as $M=1$ and $m=1$. {In the gray region} (I) in the $l-s$ {space, the} test particle can have timelike circular orbits. {In the red region (II) and yellow region} (III), the test particle does not have stable timelike circular orbits. {In the region (III), the motion} of the test particle is spacelike and unphysical.}
\label{superluminal}
\end{figure*}

We have mentioned that {the MPD equations} of the spinning test particle is obtained under the ``pole-dipole'' approximation, which will lead to the four-velocity transform from timelike to spacelike if the particle spin is too large. In order to make sure the motion of the spinning test particle is timelike, we adopt the superluminal constraint \cite{zhang2018:zwgsl}
\begin{eqnarray}
\frac{u^\mu u_\mu}{(u^t)^2} =
  \frac{g_{tt}}{c^2}
  + g_{rr}\Big(\frac{\dot{r}}{c}\Big)^2
  + g_{\phi\phi}\Big(\frac{\dot{\phi}}{c}\Big)^2
  + 2g_{\phi t}\dot{\phi}<0. \label{velocitysquare}
\end{eqnarray}
By using the superluminal constraint and circular orbit conditions \eqref{condition1} and \eqref{condition2} for the spinning test particle, we obtain the parameter space $(\bar{s}-{l})$ in Fig. \ref{superluminal}, {which} describes whether the motion on a circular orbit is timelike or spacelike. {We can} compare the results in Fig. \ref{double-circular-orbits} {and in Fig.} \ref{superluminal} to check whether the motion of the spinning test particle is timelike. We confirm that the motion of {the particle with the same spin and the same} total angular momentum that can move in two separate orbits is timelike. We have known that the effects from the GB term on the motion of the test particle can be attractive or repulsive, where a positive GB coupling parameter $\alpha$ leads to an attractive force and a negative one results in a repulsive force. The spin-curvature force can also be attractive or repulsive. The same directions of the spin angular momentum and orbital angular momentum of the spinning test particle {will lead to attractive spin-curvature force}, wile the opposite direction will lead a repulsive force. When the effects induced by the GB term and spin-curvature force exist simultaneously, the total attractive or negative {force} will be enhanced or weakened. Then it will {change} the shapes of the regions in $(\bar{l}-\bar{j})$. Thus, when the GB coupling parameter $\alpha>0$, the attractive effects will {lead to disappearance of}  the parts of the top left and bottom right of the region (II). While for the case of $\alpha<0$, the repulsive effects will {result in disappearance of} the parts of the top right and bottom left of the region (II).


Next we will investigate the ISCO of the spinning test particle. The ISCO of the test particle locates at the position where the maximum and minimum of the effective potential merge. Thus, the effective potential of {the} test particle at the ISCO  should satisfy
\begin{equation}
\frac{d^2V_{\text{eff}}}{dr^2}=0\label{condition3}.
\end{equation}
By using Eqs. \eqref{condition1}, \eqref{condition2}, and \eqref{condition3}, we can derive the ISCO of the test particle. In Ref. \cite{Guo2020}, the authors showed that the radius of the ISCO for a spinless test particle varies in the form of
\begin{equation}
r_{ISCO}=6M-\frac{11}{18}\alpha+\mathcal{O}(\alpha).
\end{equation}
This result {was} derived under the linear approach with a small $\alpha$ around $0$. Obviously, the ISCO of a spinless test particle can be larger or smaller due to the existence of the GB term. This phenomenon is consistent with the {behavior} of the effective potential, see the subfigure (a) in Fig. \ref{effectivefigure}.

\begin{figure*}[!htb]
\centering
\includegraphics[width=\linewidth]{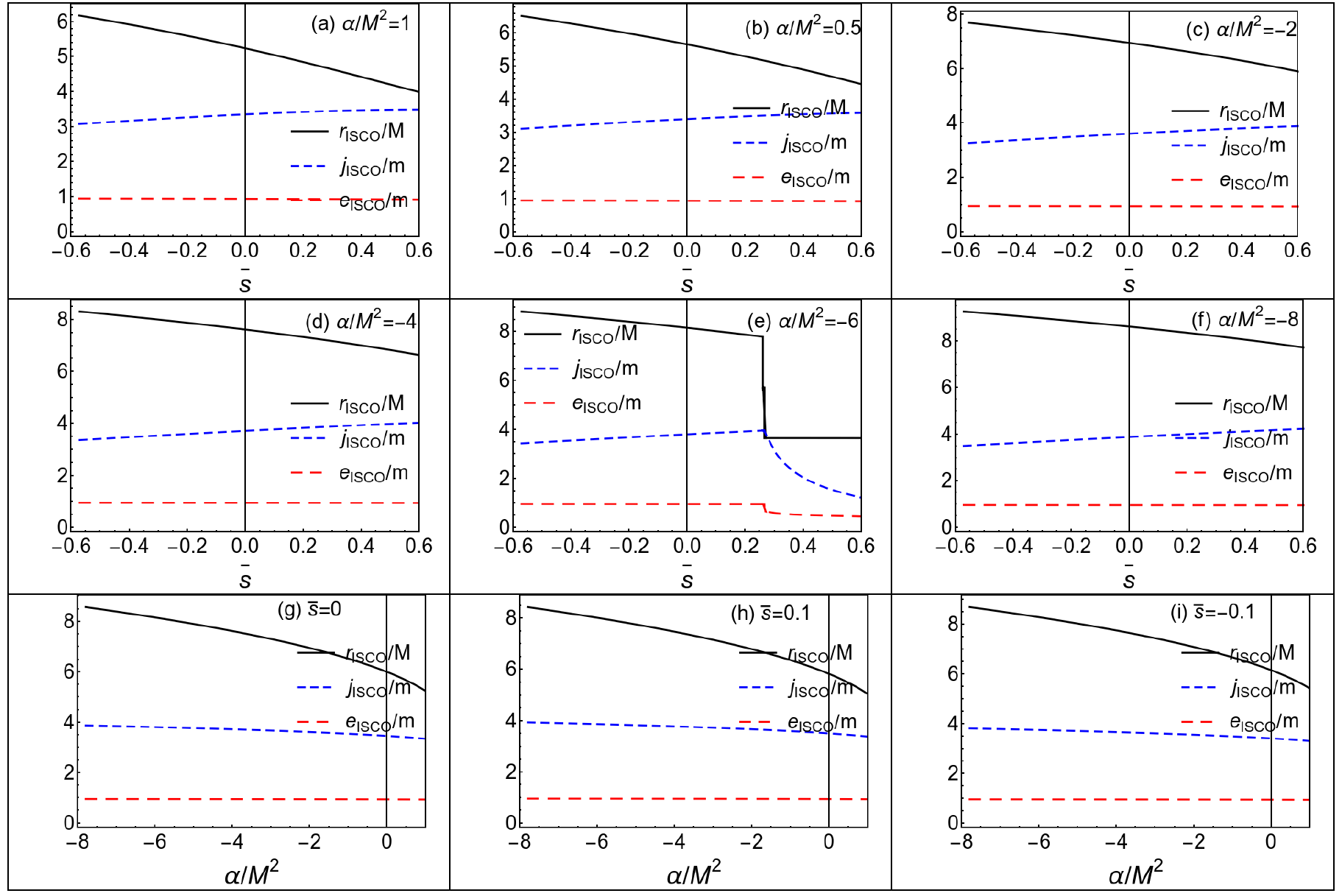}
\caption{The ISCO parameters of the spinning test particle with different values of $\alpha$. The parameters  are set as $M=1$ and $m=1$. }
\label{ISCO-figure}
\end{figure*}

When the test particle possesses a non-vanishing spin, the {contribution} of the spin-curvature force should affect the properties of the motion. The {relation} between the effective potential and {the spin} of the test particle {is still shown} in Fig. \ref{effectivefigure}. We give the numerical results of the ISCO in Fig. \ref{ISCO-figure}. Note that, there is a jump behavior for the ISCO parameters in {the subfigure} (e) in Fig. \ref{ISCO-figure}, which is induced by {the fact} that the effective potential has two minima. Because we use {the position} where the maximum and minimum of the effective potential merge to locate the ISCO and our step length of spin is not small enough to cover the change of the ISCO parameters. We summarize how the ISCO of the spinning test particle depends on the spin $\bar{s}$ and GB coupling parameter $\alpha$ as follows:
\begin{itemize}
\item For the ISCO of the spinning test particle in four-dimensional EGB black hole, the corresponding radius and angular momentum decrease with the spin $\bar{s}$ when the GB coupling parameter $\alpha$ is fixed. And when {the effect} from the GB term is considered, the spinning test particle can orbit at more smaller radius of the ISCO than the {case} of the Schwarzchild black hole in GR, and the Gauss-Bonnet term does not change the laws of the ISCO with spin.

\item When the spin of the test particle is fixed, the radius and angular momentum of the ISCO decrease with the GB coupling parameter and this behavior is almost {the same as} the results of the spinless case in Ref. \cite{Guo2020}.

\item When the spin of the spinning particle and GB coupling parameter are in the region of $\bar{s}-\bar{j}$ that the particle can have two separate orbits, the radius and angular momentum of the ISCO will become more smaller than the {case} of {the} spinless test particle in the EGB black hole \cite{Guo2020} or {the case} of {the} spinning test particle in the Schwarzchild black hole in GR \cite{zhang2018:zwgsl}.
\end{itemize}

\section{Summary and conclusion}\label{Conclusion}

In this paper, we investigated the motion of {a spinning} test particle in the equatorial plane of the four-dimensional novel EGB black hole. We solved the four-momentum and four-velocity of {the particle} and investigated how {its motion} depends on the four-dimensional GB term and particle's spin. We found that the ISCO of the spinning test particle has the similar {behavior as the case} of a spinning test particle in general relativity. And the GB term and spin parameter $\bar{s}$ can make the radii of {the} ISCO become larger or smaller. The new feature for the motion of the spinning test particle is that {it can} move at two separate orbits with {the same spin $\bar{s}$ and same} total angular momentum $\bar{j}$ when the GB coupling parameter $\alpha$ is in a special range of $-6.1<\alpha/M^2<-2$. We {also} gave the superluminal constraint on the four-velocity of the spinning test particle in {the} circular orbits and confirmed that the motion of {the spinning test particle that can move at two seperate orbits} is timelike.

\section{Acknowledgments}
This work was supported in part by the National Natural Science Foundation of
China (Grants No. 11875151, No. 11705070, No. 11522541, and No. 11675064), Y.P. Zhang was supported by the
scholarship granted by the Chinese Scholarship Council (CSC).


\end{document}